# Frequency-Phase-Locking Mechanism inside DC SQUIDs and The Analytical Expression of Current-Voltage Characteristics

Yongliang Wang [a,b*]

[a] *State Key Laboratory of Functional Materials for Informatics, Shanghai Institute of Microsystem and Information Technology (SIMIT), Chinese Academy of Sciences (CAS), Shanghai 200050, China*
[b] *CAS, Center for Excellence in Superconducting Electronics (CENSE), Shanghai 200050, China*

[*]Corresponding author. Tel.: +86 02162511070; Fax: +86 02162127493.

E-mail address: wangyl@mail.sim.ac.cn

**Abstract**

Direct-current superconducting quantum interference devices (dc SQUIDs) are ultra-sensitive flux-to-voltage convertors widely applied for biomagnetism and geophysics; they are working as the magnetic-field-effect transistors (MFETs) for their flux-modulated current-voltage characteristics. However, unlike semiconductor FETs, dc SQUIDs lack general analytical models and expressions to interpret the inner dynamics and outside current-voltage characteristics. This work presents a frequency-phase-locking (FPL) model and the analytical expression to reveal the how the current-voltage characteristics are formed inside dc SQUIDs, and how the characteristics are quantitively decided by the circuit parameters. The application of the analytical expression for the calculations of current-voltage characteristics is demonstrated; the results are compared with the numerical simulations. It is shown that a dc-SQUID is an FPL system inside and a current-modified nonlinear resistor outside; its current-voltage characteristics formed by FPL mechanism are the projections of three network impedances driven by Josephson currents. Based on the FPL model and the analytical expression, dc SQUIDs can be simply treated as nonlinear resistors in practical applications; their performance can be directly predicted and evaluated according to three network impedances.

# 1. Introduction

Direct-current Superconducting Quantum Interference Devices (dc-SQUIDs) [1] are the ultra-sensitive flux-to-voltage convertors widely applied in the subtle magnetic field measurements for biomagnetism [2] [3] and geophysical applications [4]. A dc-SQUID is a superconducting loop with two Josephson junctions connected in parallel, as shown in Fig. 1(a). It outputs a time-averaged voltage $V_s$ in response to the input flux $\Phi_i$ when it biased with a current $I_b$, and achieves the flux-modulated current-voltage characteristic in form of $F(V_s, I_b, \Phi_i) = 0$. Based on the flux-modulated characteristic, the magnetic-field-effect transistor (MFET) with current input is implemented by a dc-SQUID tightly coupled with an input coil $L_i$, as shown in Fig. 1(b); it is further working in the flux-locked loop (FLL) circuit [5] to develop magnetometers or gradiometers by connecting a pick-up coil $L_p$ to the current input, as shown in Fig. 1(c).

The flux-modulated current-voltage characteristics of dc SQUIDs are resulted from the resonances of two Josephson currents interfering inside the washer [6-9]. The washer is coupled with the pick-up and input coils in dc-SQUID magnetometers or gradiometers, as illustrated in Fig. 2. By modeling Josephson junctions as the resistively-capacitively-shunted junctions (RCSJs) [10], and considering the effects of the parasitic inductances and capacitances inside the washer and coils [11-13], the equivalent circuit of dc SQUIDs is drawn as shown in Fig. 3; it is a two-port resistor-capacitor-inductor ($RLC$) network driven by Josephson currents; its current-voltage characteristics are numerically simulated by solving differential equations [14,15], or using the simulation tools special for Josephson junction circuits [16-18].

However, numerical simulations cannot provide the analytical expression of dc SQUIDs to quantitatively describe how the current-voltage characteristics are formed by circuit elements. The analytical expression is advantageous to interpret the mechanism of SQUID-based FETs, and to implement the design evaluation directly through circuit parameters. In nearly 60 years of the history of dc SQUIDs, only a few analytical expressions are proposed [19-22]; they are valid for specific circuit models or simplified differential equations. The general analytical expression for practical dc-SQUID magnetometers or gradiometers is still absent.

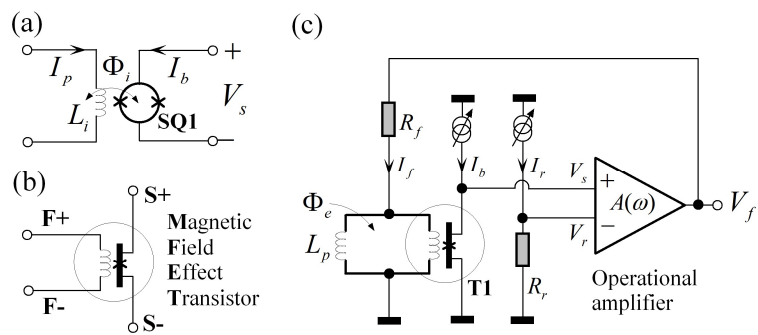

Fig. 1. Dc-SQUID and its application circuit: (a) dc-SQUID coupled with an input coil; (b)magnetic-field-effect transistor (MFET) concept; (c) flux-locked loop (FLL) circuit based on the MFET.

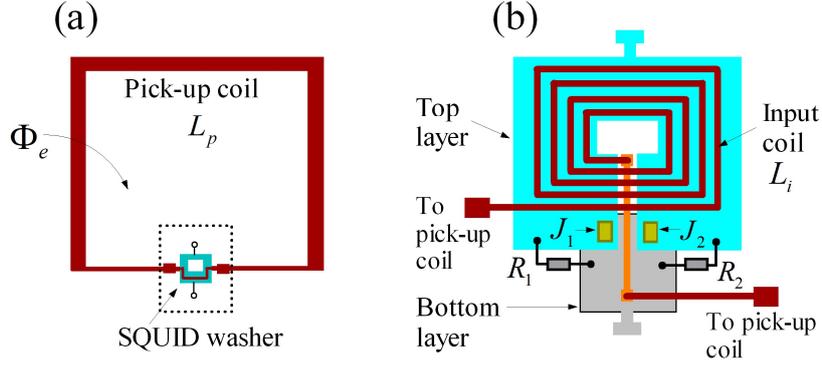

Fig. 2. (a) Layout of the niobium-based dc-SQUID magnetometer and (b) the SQUID washer, where Josephson junctions $J_1$ and $J_2$ are shunted by resistors $R_1$ and $R_1$. The pick-up coil is connected to the input coil above the SQUID washer to turn the external flux $\Phi_e$ into the superconducting current and generate an input flux $\Phi_i$ to the SQUID loop.

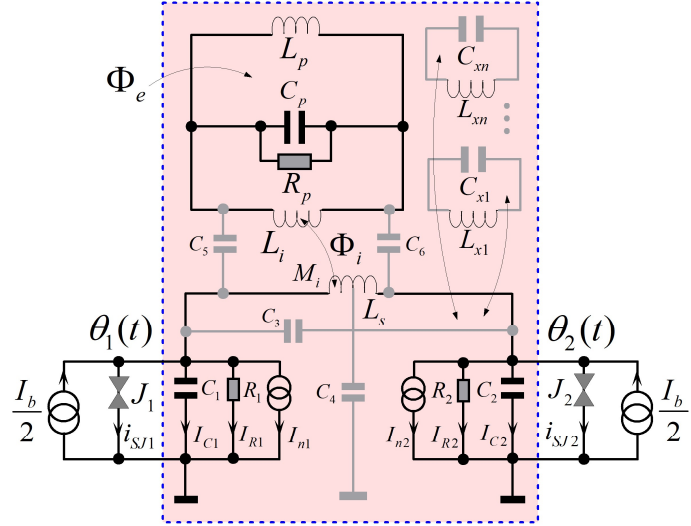

Fig. 3. Equivalent circuit of dc-SQUID magnetometers or gradiometers, where each Josephson junction is biased with a dc current of $I_b/2$. This dc SQUID circuit is exactly a two-port $RLC$ network driven by two Josephson currents: pure Josephson currents $i_{SJ1}$ and $i_{SJ2}$ are the dependent current sources of the Josephson quantum phases $\theta_1$ and $\theta_2$, namely $i_{SJ1} = I_{01}\sin\theta_1$ and $i_{SJ2} = I_{02}\sin\theta_2$; $I_{01}$ and $I_{02}$ are the critical currents; $I_{n1}$ and $I_{n2}$ are the current noises. The circuit between two Josephson currents is consisted of $RLC$ elements: $R_1$ and $R_2$ are the shunt resistors; $C_1$ and $C_2$ are the junction capacitors; $L_s$ is the inductance of SQUID washer; $M_i$ is the mutual inductance between $L_i$ and $L_s$; $L_i$ and $L_p$ are the inductances of input and pick-up coils; $C_p$ and $R_p$ are the parasitic capacitor and shunt resistor at the terminals of $L_i$. Moreover, $C_3$, $C_4$ are the parasitic capacitances inside the SQUID washer; $C_5$, $C_6$ are the ones between washer and input coil; $C_{x1}$~ $C_{xn}$ and $L_{x1}$~ $L_{xn}$ are the parasitic capacitances and inductances that arouse LC resonances inside superconducting loops.

In this article, we present a frequency-phase-locking (FPL) model to interpret the interference mechanism inside dc SQUIDs, and derive a general analytical expression of the flux-modulated current-voltage characteristics for any practical dc SQUIDs, using conventional two-port network theories. The applications of the analytical expression are demonstrated in the calculations of current-voltage characteristics for a practical dc SQUID circuit. It is revealed that a dc-SQUID circuit is working as the FPL system inside and working as a flux-modified nonlinear resistor in the FLL circuit; its current-voltage characteristics are shaped by three z parameters, namely $Z_{11}$, $Z_{22}$ and $Z_{12}$, of the linear RLC network between two Josephson junctions.

## 2. Theory

### 2.1 Two-port network analysis

Since the dc-SQUID equivalent circuit shown in the dashed box in Fig. 3, is a linear two-port RLC network, the general circuit model for dc SQUIDs is drawn, as shown in Fig. 4, where Josephson currents $i_{SJ1}(t)$ and $i_{SJ2}(t)$ under a non-zero $V_s$ will impose both dc and ac currents to the RLC network. Accordingly, $i_{SJ1}(t)$ and $i_{SJ2}(t)$ can be written with Fourier series as

$$i_{SJ1,2}(t) = i_{dc1,2} + i_{ac1,2}(t)$$
$$i_{ac1,2}(t) = \sum_{k=1}^{\infty} \text{Re}\left(I_{ac1,2}(k) \cdot e^{jk\omega_{rf}t}\right); I_{ac1,2}(k) = |I_{ac1,2}(k)| \cdot e^{j\delta_{1,2}(k)} \quad (1)$$

where $i_{dc1}$ and $i_{dc2}$ are the dc components; $i_{ac1}(t)$ and $i_{ac2}(t)$ are the ac components; $\omega_{rf}$ is the fundamental resonance frequency; $I_{ac1}(k)$ and $I_{ac2}(k)$ are the harmonic phasors defined for sinusoidal steady-state analyses [23]; Re () is the operation of getting the real part of phasors.

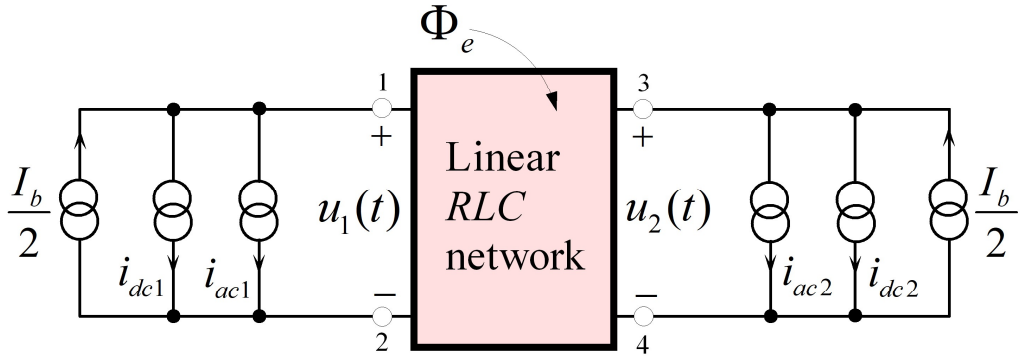

Fig. 4. Two-port network model of dc-SQUID circuits, where two Josephson currents are driving the two-port linear RLC network connected between them.

In response to the Josephson currents, the voltages $u_1(t)$ and $u_2(t)$ at two ports can also be expanded with Fourier series as

$$u_{1,2}(t) = V_s + u_{ac1,2}(t)$$
$$u_{ac1,2}(t) = \sum_{k=1}^{\infty} \text{Re}\left(V_{ac1,2}(k) \cdot e^{jk\omega_{rf}t}\right) \quad (2)$$

where the $V_s$ is the dc components at two ports; $u_{ac1}(t)$ and $u_{ac2}(t)$ are the ac components; $V_{ac1}(k)$ and $v_{ac2}(k)$ are the phasors of the harmonics.

The transfer function of the two-port network is defined with $z$ parameters, namely $Z_{11}(k\omega_{rf})$, $Z_{12}(k\omega_{rf})$, $Z_{21}(k\omega_{rf})$ and $Z_{22}(k\omega_{rf})$. Considering that the two-port network between two Josephson currents is a reciprocal circuit, $Z_{12}(k\omega_{rf}) = Z_{21}(k\omega_{rf})$, the current-to-voltage transfer function of the two-port network is written as

$$\begin{bmatrix} V_{ac1}(k) \\ V_{ac2}(k) \end{bmatrix} = \begin{bmatrix} Z_{11}(k\omega_{rf}) & Z_{12}(k\omega_{rf}) \\ Z_{12}(k\omega_{rf}) & Z_{22}(k\omega_{rf}) \end{bmatrix} \cdot \begin{bmatrix} -I_{ac1}(k) \\ -I_{ac2}(k) \end{bmatrix} = \begin{bmatrix} |V_{ac1}(k)| \cdot e^{j\sigma_1(k)} \cdot e^{j(\delta_1(k)+\pi)} \\ |V_{ac2}(k)| \cdot e^{j\sigma_2(k)} \cdot e^{j(\delta_2(k)+\pi)} \end{bmatrix}$$

$$\begin{cases} |V_{ac1}(k)| = \left| |I_{ac1}(k)| Z_{11}(k\omega_{rf}) + |I_{ac2}(k)| Z_{12}(k\omega_{rf}) \cdot e^{j\Delta\delta(k)} \right| \\ |V_{ac2}(k)| = \left| |I_{ac2}(k)| Z_{22}(k\omega_{rf}) + |I_{ac1}(k)| Z_{12}(k\omega_{rf}) \cdot e^{-j\Delta\delta(k)} \right| \\ e^{j\sigma_1(k)} = \left( |I_{ac1}(k)| Z_{11}(k\omega_{rf}) + |I_{ac2}(k)| Z_{12}(k\omega_{rf}) \cdot e^{j\Delta\delta(k)} \right) / |V_{ac1}(k)| \\ e^{j\sigma_2(k)} = \left( |I_{ac2}(k)| Z_{22}(k\omega_{rf}) + |I_{ac1}(k)| Z_{12}(k\omega_{rf}) \cdot e^{-j\Delta\delta(k)} \right) / |V_{ac2}(k)| \\ \Delta\delta(k) = \delta_2(k) - \delta_1(k) \end{cases} \quad (3)$$

where $\sigma_1(k)$ and $\sigma_2(k)$ are the extra phases shifted by network impedances; $\delta(k)$ is the phase difference between $I_{ac1}(k)$ and $I_{ac2}(k)$.

For a given two-port network, the $z$ parameters can be derived from the equivalent circuit, as demonstrated in Appendix-I, and can be also directly extracted from the circuit layout using electronic-design-automation (EDA) tools [24].

In the dc state with $\omega_{rf}= 0$, $Z_{11}(0) = Z_{22}(0) = Z_{21}(0)$; the dc voltage $V_s$ is generated by the dc current flowing through $Z_{11}(0)$, thus

$$\begin{aligned} V_s &= Z_{11}(0) \cdot (I_b - i_{cmm}) \\ i_{cmm} &= i_{dc1} + i_{dc2} \end{aligned} \quad (4)$$

Here $i_{cmm}$ is the total dc current bypassed by two Josephson currents.

## 2.2 Flux-quantization law

Through the integrals of $u_1(t)$ and $u_2(t)$ with respect to time, we can derive the Josephson phases, $\theta_1(t)$ and $\theta_2(t)$, in form of Fourier series as

$$\theta_{1,2}(t) = \theta_{dc1,2} + \frac{2\pi V_s}{\Phi_0} t + \theta_{ac1,2}(t)$$

$$\theta_{ac1,2}(t) = \sum_{k=1}^{\infty} \text{Re}\left( \frac{2\pi}{\Phi_0} \cdot \frac{|V_{ac1,2}(k)|}{jk\omega_{rf}} \cdot e^{j\sigma_{1,2}(k)} \cdot e^{j(\delta_{1,2}(k)+\pi)} \cdot e^{jk\omega_{rf}t} \right) \quad (5)$$

$$= \sum_{k=1}^{\infty} \frac{2\pi}{\Phi_0} \cdot \frac{|V_{ac1,2}(k)|}{k\omega_{rf}} \cdot \sin(k\omega_{rf}t + \sigma_{1,2}(k) + \delta_{1,2}(k) + \pi)$$

where $\theta_{dc1}$ and $\theta_{dc2}$ are the dc components; $\theta_{ac1}(t)$ and $\theta_{ac2}(t)$ are the ac components; $\Phi_0 = h/2e = 2.07 \times 10^{-15}$ Wb is flux quantum.

According to the flux-quantization law [25], the difference $\Delta\theta_{dc}$ between $\theta_{dc1}$ and $\theta_{dc2}$ is created by the input flux $\Phi_i$ and the flux induced inside the SQUID loop, namely

$$\Delta\theta_{dc} = \theta_{dc2} - \theta_{dc1} = \frac{2\pi}{\Phi_0} \cdot \left(\Phi_i - L_s^* i_{cir}\right)$$

$$i_{cir} = \frac{i_{dc2} - i_{dc1}}{2} \tag{6}$$

where $i_{cir}$ is the differential current of two Josephson currents, and $L_s^*$ is the overall inductance of the SQUID loop. The $L_s^*$ is modified by the pick-up and input coils; its derivation can be found in Appendix-II.

## 2.3 Dc Josephson effect

The time-averaged active powers of $i_{SJ1}(t)$ and $i_{SJ2}(t)$, $P_{SJ1}$ and $P_{SJ2}$, are defined as

$$P_{SJ1,2} = \frac{1}{T}\int_0^T i_{SJ1,2}(t) u_{1,2}(t) dt = \frac{\Phi_0}{2\pi} \cdot \left.\frac{I_{01,2}\cos\theta_{1,2}(t)}{T}\right|_0^T \approx 0; \left(T \gg 2\pi/\omega_{rf}\right) \tag{7}$$

They are close to zero for the period $T$ ($T \gg 2\pi/\omega_{rf}$), which depicts the non-loss principle of Josephson currents. Since $P_{SJ1}$ and $P_{SJ2}$ can be also written with phasors of currents and voltages as

$$P_{SJ1,2} = V_s \cdot i_{dc1,2} + \sum_{k=1}^{\infty} \text{Re}\left(\frac{V_{ac1,2}(k) \cdot \overline{I_{ac1,2}(k)}}{2}\right) \tag{8}$$

where the bar over $I_{ac1,2}(k)$ indicates the conjugate operation, we can calculate the dc Josephson currents with the ac components as

$$\begin{cases} i_{dc1} = \sum_{k=1}^{\infty} i_{dc1\_k}; i_{dc2} = \sum_{k=1}^{\infty} i_{dc2\_k} \\ i_{dc1\_k} = \text{Re}\left(\frac{V_{ac1}(k) \cdot \overline{I_{ac1}(k)}}{-2V_s}\right) \\ i_{dc2\_k} = \text{Re}\left(\frac{V_{ac2}(k) \cdot \overline{I_{ac2}(k)}}{-2V_s}\right) \end{cases} \tag{9}$$

where $i_{dc1\_k}$ and $i_{dc2\_k}$ are the dc components mixed from $V_{ac1}(k)$ and $v_{ac2}(k)$ by $I_{ac1}(k)$ and $I_{ac2}(k)$, respectively. The balance between ac and dc power consumptions makes Josephson currents work as the mixers to generate dc currents.

## 2.4 Ac Josephson effect

Using the Fourier series of Josephson phases in (5), we rewrite the Josephson currents $i_{SJ1}(t)$ and $i_{SJ2}(t)$ as

$$i_{SJ1,2}(t) = I_{01,2}\sin\left(\theta_{dc1,2} + \frac{2\pi V_s}{\Phi_0}t + \theta_{ac1,2}(t)\right)$$

$$= I_{01,2}\text{Re}\left(e^{j(\theta_{dc1,2}-0.5\pi)} \cdot e^{j\frac{2\pi V_s}{\Phi_0}t} \cdot \prod_{k=1}^{\infty} e^{j\frac{2\pi}{\Phi_0}\cdot\frac{|V_{ac1,2}(k)|}{k\omega_{rf}}\cdot\sin(k\omega_{rf}t+\sigma_{1,2}(k)+\delta_{1,2}(k)+\pi)}\right) \tag{10}$$

where the ac components introduced by $\theta_{ac1}(t)$ and $\theta_{ac2}(t)$ are in the form of $e^{jx\sin y}$ which can be further expanded with Bessel functions.

Therefore, the phasors $I_{ac1,2}(k)$ defined in (1) can be finally expressed with Bessel functions, and the ac Josephson currents are working as a group of voltage-controlled-oscillators (VCOs); their fundamental frequency $\omega_{rf}$ is decided by $V_s$ as

$$\omega_{rf} = \frac{2\pi}{\Phi_0} \cdot V_s \tag{11}$$

### 2.5 Frequency-phase-locking mechanism

By synthesizing all the circuit equations from (1) to (11), the interference mechanism inside dc SQUIDs is analytically depicted with a general FPL model, as shown in Fig. 5. The working principle of dc SQUIDs is interpreted as follows:

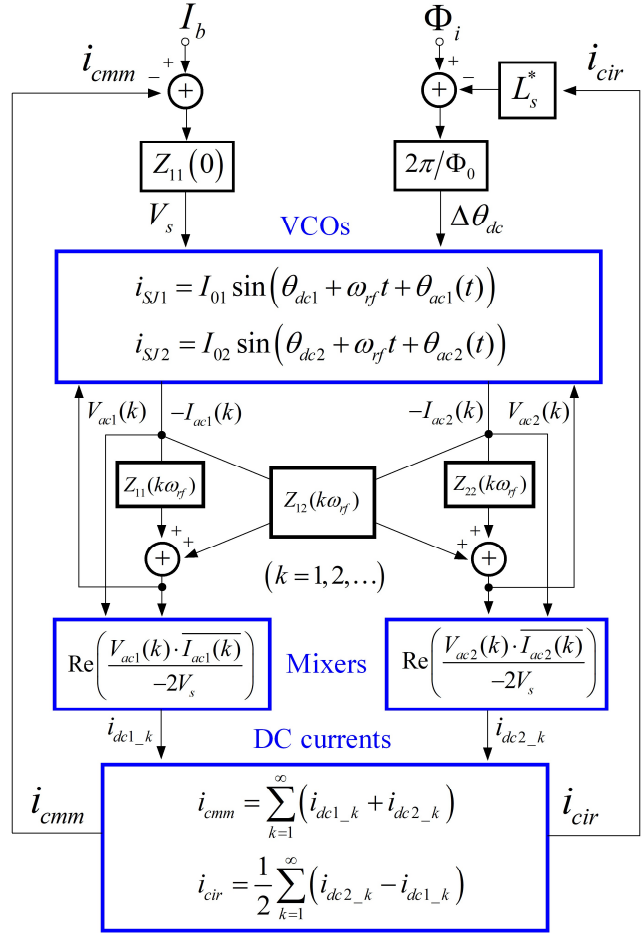

Fig. 5. FPL model of dc SQUIDs, where Josephson currents are functioning as both the VCOs and mixers to extract the z parameters of the two-port network; to the given input $I_b$ and $\Phi_i$, the FPL system will be locked at a stable $\omega_{rf}$ and $\Delta\theta_{dc}$ that is how the dc-SQUID achieves the flux-modulated current-voltage characteristics.

(1) Each Josephson current works as both VCOs and mixers. The VCOs impose a group of harmonic ac currents to the two-port $RLC$ network, while the mixers turn the voltage responses at two ports into dc currents.

(2) Three impedances of the two-port $RLC$ network implements the current-to-voltage transfer between VCOs and mixers. The impedance $Z_{12}(k\omega_{rf})$ (k =1, 2, ...)

induces the interference between two Josephson currents.

(3) The dc current $i_{cmm}$ is fed back to adjust the $V_s$ and alter the frequency $\omega_{rf}$, while $i_{cir}$ is sent back to adjust the $\Delta\theta_{dc}$. The FPL system outputs a static $V_s$ when it is locked at a stable $\omega_{rf}$ and $\Delta\theta_{dc}$ to the given $I_b$ and $\Phi_i$; this is how the flux-modulated current-voltage characteristics of dc SQUIDs are formed.

## 2.6 Analytical expression of current-voltage characteristics

The FPL model contains two analytical expressions to implement the function $F(V_s, I_b, \Phi_i) = 0$; we define them as $\Phi_i = f_\Phi(V_s, \Delta\theta_{dc})$ and $I_b = f_I(V_s, \Delta\theta_{dc})$ as

$$F(V_s, I_b, \Phi_i) = 0 \Rightarrow \begin{cases} \Phi_i = f_\Phi(V_s, \Delta\theta_{dc}) = \dfrac{\Phi_0}{2\pi} \cdot \Delta\theta_{dc} + \Phi_{cir} \\ I_b = f_I(V_s, \Delta\theta_{dc}) = \left(\dfrac{1}{Z_{11}(0)} + G_{cmm}\right) \cdot V_s \end{cases} \tag{12}$$

where $\Phi_{cir}$ is the flux generated by $i_{cir}$ and $G_{cmm}$ is the conductance created by $i_{cmm}$.

The expression of $\Phi_{cir}$ is expanded with the network impedances as

$$\begin{aligned} &\Phi_{cir} = L_s^* i_{cir} = \Phi_{cir1} + \Phi_{cir2} \\ &\begin{cases} \Phi_{cir1} = \dfrac{L_s^*}{4V_s} \sum_{k=1}^{\infty} \mathrm{Re}\left(|I_{ac2}(k)|^2 Z_{22}(k\omega_{rf}) - |I_{ac1}(k)|^2 Z_{11}(k\omega_{rf})\right) \\ \Phi_{cir2} = \dfrac{L_s^*}{2V_s} \sum_{k=1}^{\infty} |I_{ac1}(k)I_{ac2}(k)| \cdot \mathrm{Im}\left(Z_{12}(k\omega_{rf})\right) \cdot \sin\Delta\delta(k) \end{cases} \end{aligned} \tag{13}$$

where Im() is the operation of getting imaginary value. In symmetric dc SQUIDs, $Z_{11}(k\omega_{rf}) = Z_{22}(k\omega_{rf})$, $\Phi_{cir2}$ is major part, and $\Phi_{cir1}$ is negligible.

The conductance $G_{cmm}$ is expressed with network impedances as

$$\begin{aligned} &G_{cmm} = \dfrac{i_{cmm}}{V_s} = G_{cmm1} + G_{cmm2} \\ &\begin{cases} G_{cmm1} = \dfrac{1}{2V_s^2} \sum_{k=1}^{\infty} \mathrm{Re}\left(|I_{ac1}(k)|^2 Z_{11}(k\omega_{rf}) + |I_{ac2}(k)|^2 Z_{22}(k\omega_{rf})\right) \\ G_{cmm2} = \dfrac{1}{V_s^2} \sum_{k=1}^{\infty} |I_{ac1}(k)I_{ac2}(k)| \cdot \mathrm{Re}\left(Z_{12}(k\omega_{rf})\right) \cdot \cos\Delta\delta(k) \end{cases} \end{aligned} \tag{14}$$

where $G_{cmm1}$ is the part associated with $Z_{11}(k\omega_{rf})$ and $Z_{22}(k\omega_{rf})$, and $G_{cmm2}$ the second part decided by $Z_{12}(k\omega_{rf})$.

In (13) and (14), the $|I_{ac1,2}(k)|$ and $\Delta\delta(k)$ are also the functions of $V_s$ and $\Delta\theta_{dc}$. Their expressions can be found as by expanding (10) using the first-class Bessel functions. In Appendix-III, we present a group of simple approximations of $|I_{ac1,2}(k)|$ and $\Delta\delta(k)$ as

$$\begin{aligned} &|I_{ac1,2}(k)| \approx J_{k-1}(h_{1,2})I_{01,2}; \Delta\delta(k) \approx k\Delta\theta_{dc} + (k-1)\Delta\sigma \\ &\begin{cases} \Delta\sigma = \sigma_2(1) - \sigma_1(1) \\ e^{j\Delta\sigma} = \dfrac{h_1\left(J_0(h_2)I_{02}Z_{22}(\omega_{rf}) + J_0(h_1)I_{01}Z_{12}(\omega_{rf}) \cdot e^{-j\Delta\theta_{dc}}\right)}{h_2\left(J_0(h_1)I_{01}Z_{11}(\omega_{rf}) + J_0(h_2)I_{02}Z_{12}(\omega_{rf}) \cdot e^{j\Delta\theta_{dc}}\right)} \end{cases} \end{aligned} \tag{15}$$

where $h_1$ and $h_2$ are two factors defined in (20).

The curve of $f_\Phi$ ($V_s$, $\Delta\theta_{dc}$) is illustrated in Fig. 6(a). It is shown that the input flux $\Phi_i$ alters the phase difference $\Delta\theta_{dc}$ with a reluctant flux $\Phi_{cir}$; the $\Delta\theta_{dc}$ maintains a monotone relation with $\Phi_i$, since the $\Phi_{cir}$ is bounded.

The characteristic of $f_I$ ($V_s$, $\Delta\theta_{dc}$) is illustrated in Fig. 6(b), where the current-voltage curve depicts the nonlinear $G_{cmm}$ created by dc Josephson currents. In (14), the $G_{cmm}$ is inversely proportional to $V_s^2$, and its component $G_{cmm2}$ is proportional to the cosine function of $\Delta\theta_{dc}$; this explains why the curves of $f_I$ ($V_s$, $\Delta\theta_{dc}$) are concave upward and flux-modulated. Meanwhile, the flux-modulation depth of $G_{cmm2}$ is decided by Re($Z_{12}(k\omega_{rf})$), especially the Re($Z_{12}(\omega_{rf})$) for the fundamental harmonic.

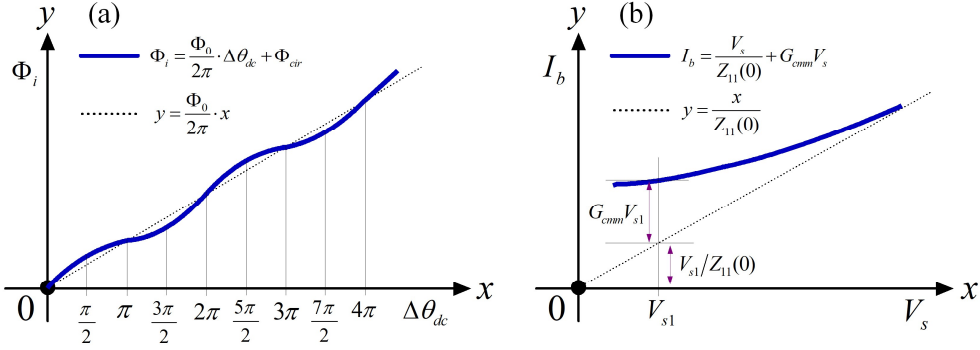

Fig. 6. (a) Curve of $\Phi_i = f_\Phi$ ($V_s$, $\Delta\theta_{dc}$); (b) curve of $I_b = f_I$ ($V_s$, $\Delta\theta_{dc}$).

## *2.7 Equivalent circuit of SQUID-based FETs*

Based on the concept of $G_{cmm}$, the equivalent circuit of SQUID-based FETs in the FLL circuit is drawn as shown in Fig. 7, where the dc-effect of Josephson currents is simulated as two nonlinear resistors connected in parallel with $Z_{11}(0)$. The conductance $G_{cmm2}$ the reason that dc SQUIDs behave as the flux-modulated nonlinear resistor.

Therefore, as semiconductor FETs are the voltage-modified nonlinear resistors, SQUID-based MFETs are the current-modified nonlinear resistors in the application circuits. This equivalent circuit can be easily understood by electronics engineers who are familiar with the circuit design of semiconductor FETs.

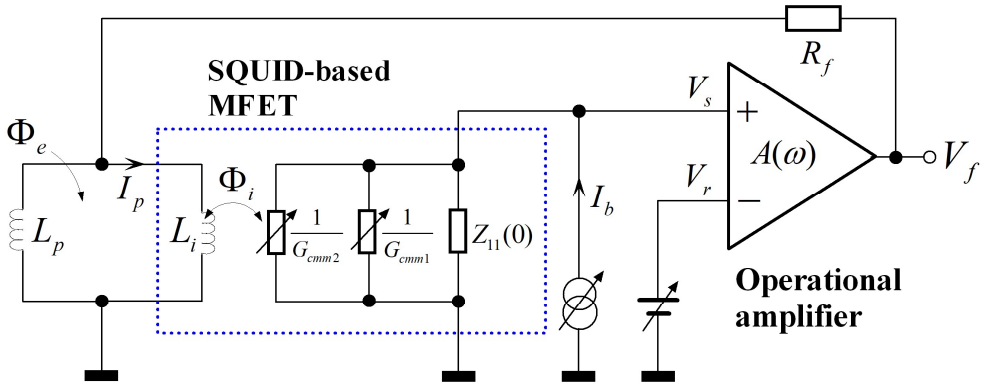

Fig. 7. Equivalent circuit of SQUID-based MFETs in the FLL circuit.

# 3. Comparison of Analytical calculations with Numerical Simulations

## 2.1 Cases and results

For the dc-SQUID circuit shown in Fig. 3, We use both the analytical and numerical methods to calculate the current-voltage characteristics; the results and the absolute value of $Re(Z_{12}(\omega_{rf}))$ are compared as shown in Fig. 8-11, where each current-voltage curve is calculated for a given $\Phi_i$; $\Phi_i$ is varied from 0 to $0.5\Phi_0$ with a step of $0.1\Phi_0$.

The analytical method firstly calculates $\Phi_i$ and $I_b$ for the given $V_s$ and $\Delta\theta_{dc}$, using $f_\Phi$ ($V_s$, $\Delta\theta_{dc}$) and $f_I$ ($V_s$, $\Delta\theta_{dc}$) in (12), then extracts the current-voltage curves from the results under the same $\Phi_i$. In the application of the analytical expression, 20 harmonics are enough to be considered ($k \leq 20$).

The numerical method finds numerical solutions of the differential circuit equation to the given $\Phi_i$ and $I_b$, then extracts the current-voltage characteristics by averaging the real-time voltages of Josephson junctions. The circuit equation of the equivalent circuit is derived according to the flux-quantization law and Kirchhoff's laws [14].

Four cases are demonstrated for the application of the analytical expression. Each case is configured with different circuit parameters. The basic circuit parameters are listed in Table 1; the rests are noted in the results. Those circuit parameters are normalized with $R_0$ and $I_0$ which are the average shunt resistance and the average critical current, respectively. For instance, any an inductance $L_x$ is normalized as $\beta_{Lx}$, $\beta_{Lx} = 2\pi I_0 L_x/\Phi_0$; a capacitance $C_x$ is normalized as $\beta_{Cx}$, $\beta_{Cx} = 2\pi I_0 R_0^2 C_x/\Phi_0$; a resistance $R_x$ is normalized as $\gamma_x$, $\gamma_x = R_x/R_0$.

Table 1. Basic parameters of the dc-SQUID circuit.

| Parameter | Symbol | Value(Unit) |
|---|---|---|
| Critical currents | $I_{01}/I_0$, $I_{02}/I_0$ | 1 |
| Shunted resistors | $\gamma_1$, $\gamma_2$ | 1 |
| Inductance of washer | $\beta_{Ls}$ | 2.2 |
| Inductance $L_i$ | $\beta_{Li}$ | 100 |
| Inductance $L_p$ | $\beta_{Lp}$ | 500 |

Case 1: the $\beta_{Mi}$ is set as 0.1 to simulate the characteristics of the bare SQUID neglecting the coupling of input loop. There is one zero-point of $Re(Z_{12}(\omega_{rf}))$ and one intersection of the current-voltage curves, as shown in Fig. 8.

Case 2: the $\beta_{Mi}$ is set as 10; three zero points of $Re(Z_{12}(\omega_{rf}))$ create three intersections in current-voltage curves, as shown in Fig. 9.

Case 3: the $\beta_{Cp}$ is increased to 0.008; the relocation of intersections of current-voltage curves are simulated, as shown in Fig. 10.

Case 4: the shunt resistor $\gamma_p$ is set as 10; it eliminates two zero points of $Re(Z_{12}(\omega_{rf}))$ and smooth the current-voltage characteristics, as shown in Fig. 11.

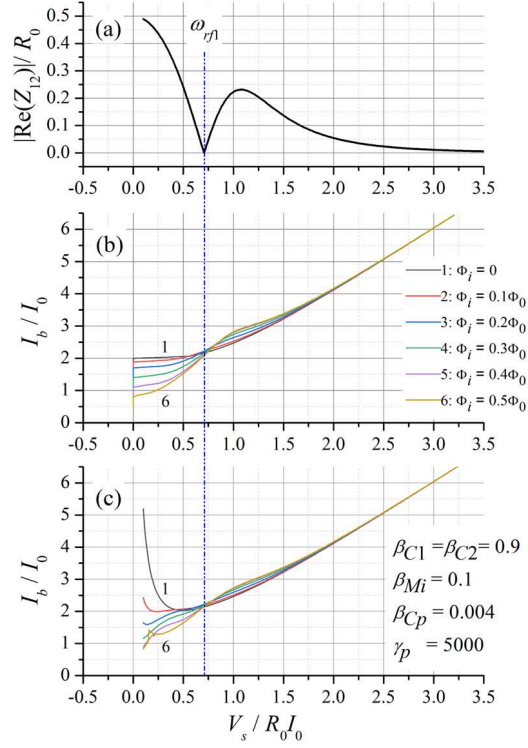

Fig. 8. Current-voltage characteristics calculated for Case 1: (a) Absolute value of Re($Z_{12}(\omega_{rf})$) versus voltage output $V_s$, where there is one zero-point at the frequency $\omega_{rf1}$; (b) current-voltage curves by numerical simulations; (c) current-voltage curves calculated with the analytical expression.

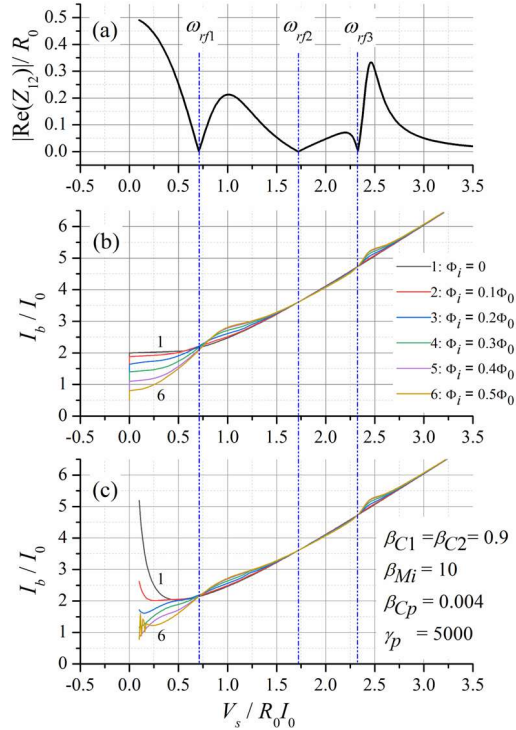

Fig. 9. Current-voltage characteristics calculated for Case 2: (a) Absolute value of Re($Z_{12}(\omega_{rf})$) versus voltage output $V_s$, where there are three zero points at frequencies $\omega_{rf1} \sim \omega_{rf3}$; (b) current-voltage curves numerically simulated; (c) current-voltage curves calculated with the analytical expression.

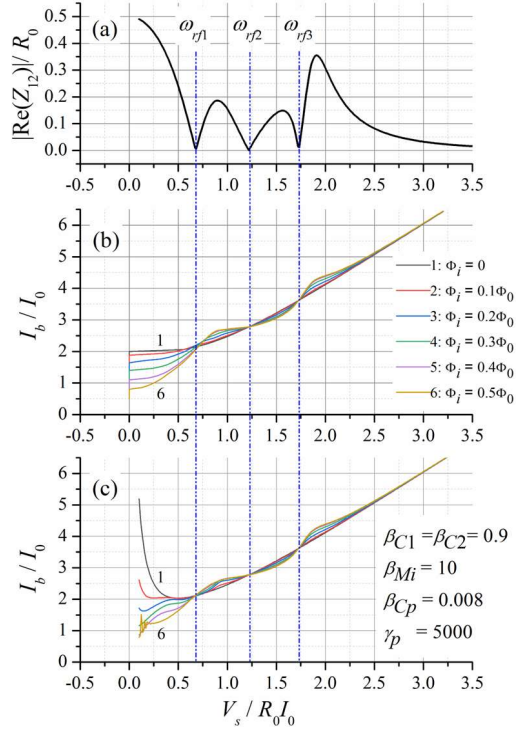

Fig. 10. Current-voltage characteristics calculated for Case 3: (a) Absolute value of Re($Z_{12}(\omega_{rf})$) versus voltage output $V_s$, where there are three zero points at frequencies $\omega_{rf1} \sim \omega_{rf3}$; (b) current-voltage curves through numerical simulations; (c) current-voltage curves calculated with the analytical expression.

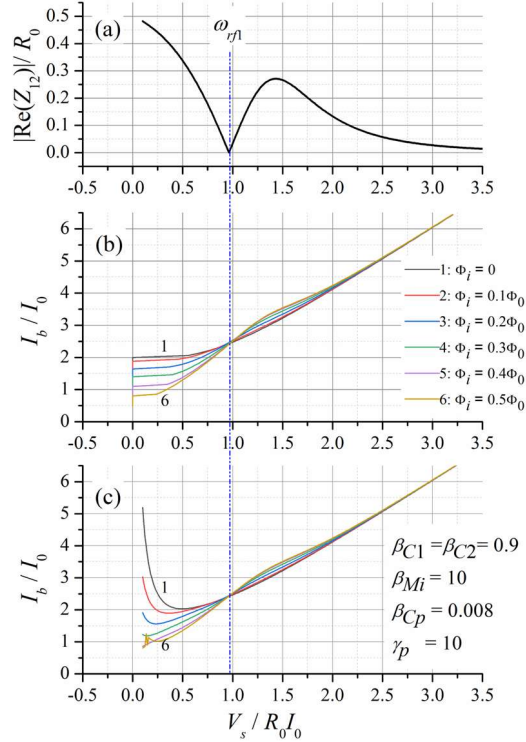

Fig. 11. Current-voltage characteristics calculated for Case 4: (a) Absolute value of Re($Z_{12}(\omega_{rf})$) versus voltage output $V_s$, which has only one zero-point at the frequency $\omega_{rf1}$; (b) current-voltage curves through numerical simulations; (c) current-voltage curves calculated with the analytical expression.

*2.2 Discussion*

Those current-voltage characteristics in four cases reproduce the typical results measured from practical dc SQUID magnetometers or gradiometers [26-28]. They are also well repeated by the analytical calculations for $V_s$ larger than $0.5R_0I_0$.

In the region with $V_s$ less than $0.5R_0I_0$, the mismatch between the analytical and numerical results is caused by the error of the approximate expressions of $I_{ac1}(k)$ and $I_{ac2}(k)$ in (15). This error increases with the decease of $V_s$, because the approximations used in (19) neglect the higher-order harmonics of $\theta_{ac1}(t)$ and $\theta_{ac2}(t)$ which will be significant for small $V_s$. The error of analytical calculations can be further reduced by finding more complex expressions of $I_{ac1}(k)$ and $I_{ac2}(k)$ from (10).

Both the numerical and analytical results demonstrate that the flux-modulation depth of current-voltage characteristics depends on the value of $Re(Z_{12}(\omega_{rf}))$, and there will be an intersection point in current-voltage curves at the zero-point of $Re(Z_{12}(\omega_{rf}))$. Accordingly, as illustrated in Fig. 9 and 10, the current-voltage characteristics of a dc-SQUID will be distorted by $Z_{12}(\omega_{rf})$ with multiple zero points.

Therefore, the impedance $Z_{12}(\omega_{rf})$ is the key parameter to evaluate the performance of dc SQUIDs for design optimizations. The zero points of $Z_{12}(\omega_{rf})$ are created by the loop inductance and the junction capacitance inside SQUID washer, the inductances of input and pick-up coils, as well as the parasitic capacitances between the washer and the coupled coils. In practical dc-SQUID design, we should reduce the zero points of $Re(Z_{12}(\omega_{rf}))$ and keep the zero-point away from working points to smooth the current-voltage characteristics.

# 4. Conclusion

We presented a FPL model and its analytical expression to interpret the inside dynamics and the outside current-voltage characteristics of dc SQUIDs through two-port network analyses. The FPL model intuitively depicts how the current-voltage characteristics are formed by Josephson currents; the analytical expression quantitively describe how network impedances shape those current-voltage characteristics; The theoretical analyses reveal that:

(1) A dc-SQUID is a two-port *RLC* network driven by two Josephson currents; its inside dynamics is depicted with the FPL mechanism; its outside behavior is the flux-modified nonlinear resistor.

(2) The equivalent resistance of the dc-SQUID is the projection of three impedances of the two-port network; the impedance $Z_{12}$ is the key parameter that induces the magnetic-flux-effect, and the $Re(Z_{12})$ decides the flux-modulation depth.

Based on those understandings, electronics engineers can treat SQUID-based MFETs simply as current-modified resistors in the magnetometer circuit design, and evaluate the current-voltage characteristics of a dc-SQUID directly through the network impedances extracted from the circuit layout, independent of either the equivalent circuit or the circuit equations.

# Appendix

## I. Two-port network impedances

For the dc-SQUID circuit shown in Fig. 3, its sinusoidal steady-state equivalent circuit is redrawn by turning inductances and capacitances into complex impedances, as shown in Fig. 12, where only major circuit elements, such as loop inductance, input and pick-up coils, are considered for simplicity.

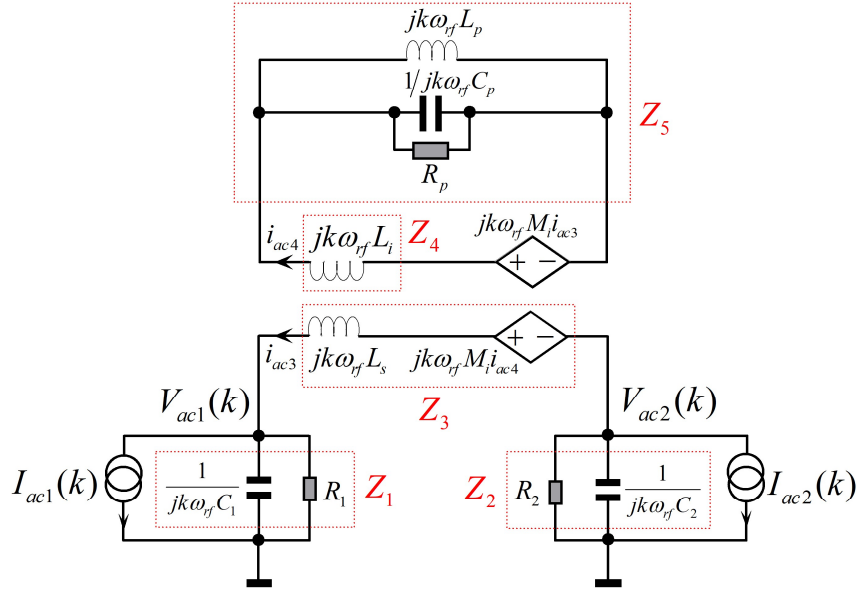

Fig. 12. Sinusoidal steady-state equivalent circuit of the dc-SQUID circuit shown in Fig. 3.

We can define the five impedances shown in Fig. 12, namely $Z_1 \sim Z_5$, as

$$Z_1 = \frac{1}{1/R_1 + jk\omega_{rf}C_1}; Z_2 = \frac{1}{1/R_2 + jk\omega_{rf}C_2}; Z_3 = jk\omega_{rf}L_s + \frac{(k\omega_{rf}M_i)^2}{Z_4 + Z_5}$$

$$Z_4 = jk\omega_{rf}L_i; Z_5 = \frac{1}{1/R_p + j(k\omega_{rf}C_p - 1/(k\omega_{rf}L_p))}$$
(16)

Since the two-port network between Josephson currents is a $\pi$-type network consisted of $Z_1$, $Z_2$ and $Z_3$, three network impedances are written as

$$Z_{11}(k\omega_{rf}) = \frac{Z_1(Z_2+Z_3)}{Z_1+Z_2+Z_3}; Z_{22}(k\omega_{rf}) = \frac{Z_2(Z_1+Z_3)}{Z_1+Z_2+Z_3}; Z_{12}(k\omega_{rf}) = \frac{Z_1Z_2}{Z_1+Z_2+Z_3}$$
(17)

## II. Effective loop-inductance

The dc-state equivalent circuit for the dc-SQUID circuit is redrawn by neglecting all the capacitances, as shown in Fig. 13. The pick-up and input coils form a superconducting loop and apply the magnetic-field screen effect to the dc-SQUID loop; thus, the effective loop inductance $L^*_s$ of the dc-SQUID is derived as

$$L^*_s = L_s - M_i^2 / (L_p + L_i)$$
(18)

It is shown that the $L^*_s$ will be reduced by both pick-up and input coils.

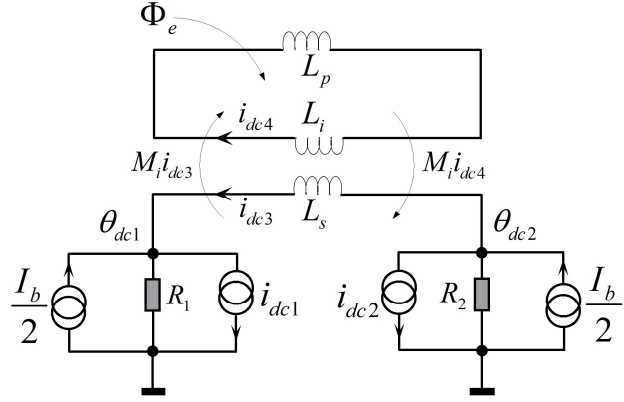

Fig. 13. Dc-state equivalent circuit of the dc-SQUID circuit shown in Fig. 3, where circuit elements are driven by the dc components of Josephson currents.

## III. Harmonics of Josephson currents

In (10), if only the component in form of $e^{jx\sin y}$ with $k=1$ is considered and expanded with the first-class Bessel functions, we can obtain the expressions of $i_{SJ1}(t)$ and $i_{SJ2}(t)$ in form of Fourier series as

$$\begin{aligned} i_{SJ1,2}(t) &= I_{01,2}\sin\left(\theta_{dc1,2}+\omega_{rf}t+\theta_{ac1,2}(t)\right) \\ &\approx I_{01,2}\mathrm{Re}\left(e^{j(\theta_{dc1,2}-0.5\pi)}\cdot e^{j\omega_{rf}t}\cdot e^{jh_{1,2}\sin(\omega_{rf}t+\sigma_{1,2}(1)+\delta_{1,2}(1)+\pi)}\right) \\ &\approx I_{01,2}\mathrm{Re}\left(e^{j(\theta_{dc1,2}-0.5\pi)}\cdot \sum_{n=1}^{\infty}J_{n-1}(h_{1,2})\cdot e^{j(n-1)(\sigma_{1,2}(1)+\delta_{1,2}(1)+\pi)}\cdot e^{jn\omega_{rf}t}\right)\end{aligned} \quad (19)$$

$$h_{1,2} = \frac{|V_{ac1,2}(1)|}{V_s}$$

where $h_1$ and $h_2$ are the amplitudes of fundamental harmonics of $\theta_{ac1}(t)$ and $\theta_{ac2}(t)$.

Referring to the definition of phasors in (1), the approximate expressions of $I_{ac1}(k)$ and $I_{ac2}(k)$ are extracted from (19) as

$$I_{ac1,2}(k) = I_{01,2}J_{k-1}(h_{1,2})\cdot e^{j(k-1)(\sigma_{1,2}(1)+\delta_{1,2}(1)+\pi)}\cdot e^{j(\theta_{dc1,2}-0.5\pi)} \quad (20)$$

For the given $V_s$, $\Delta\theta_{dc}$, $h_1$ and $h_2$ can be solved from the expressions of $I_{ac1}(1)$ and $I_{ac2}(1)$, according to their definition in (19), namely

$$\begin{aligned} h_1 &= \frac{1}{V_s}\left|J_0(h_1)I_{01}Z_{11}(\omega_{rf})+J_0(h_2)I_{02}Z_{12}(\omega_{rf})\cdot e^{j\Delta\theta_{dc}}\right| \\ h_2 &= \frac{1}{V_s}\left|J_0(h_2)I_{02}Z_{22}(\omega_{rf})+J_0(h_1)I_{01}Z_{12}(\omega_{rf})\cdot e^{-j\Delta\theta_{dc}}\right| \end{aligned} \quad (21)$$

The corresponding iteration formula for solving the $h_1$ and $h_2$ are written as

$$h_{1,2} = h_{1,2}^{N}$$
$$\begin{cases} h_1^0 = 0; h_2^0 = 0 \\ h_1^n = h_1^{n-1} + \lambda \cdot \left( \left| J_0(h_1^{n-1})I_{01}Z_{11}(\omega_{rf}) + J_0(h_2^{n-1})I_{02}Z_{12}(\omega_{rf}) \cdot e^{j\Delta\theta_{dc}} \right| - h_1^{n-1}V_s \right) \\ h_2^n = h_2^{n-1} + \lambda \cdot \left( \left| J_0(h_2^{n-1})I_{02}Z_{22}(\omega_{rf}) + J_0(h_1^{n-1})I_{01}Z_{12}(\omega_{rf}) \cdot e^{-j\Delta\theta_{dc}} \right| - h_2^{n-1}V_s \right) \\ n = 1, 2, \ldots N; \left| h_{1,2}^N - h_{1,2}^{N-1} \right| < \varepsilon \end{cases} \quad (22)$$

where $\lambda$ is the iteration step and $\varepsilon$ is the calculation error; $\lambda$ is usually set less than 1 to keep the convergence of the iteration; the solution is found when the difference between two iterative results is less than $\varepsilon$.

# Acknowledgements